\documentclass[10pt,a4paper,notitlepage,prd,showpacs,preprintnumbers,superscriptaddress,nofootinbib]{revtex4-1}
\usepackage{float}
\usepackage{amsmath}
\usepackage{cancel}
\usepackage{graphicx}

\makeatletter

\providecommand{\tabularnewline}{\\}

\usepackage{bm}\usepackage{tensor}
\usepackage{comment}
\usepackage{ifpdf}
\usepackage{slashed}
\usepackage{color}
\usepackage[mathscr]{eucal}

\ifpdf
  \usepackage{graphicx}     
  \usepackage[bookmarksopen,colorlinks=true,linkcolor=bblue,citecolor=bblue,urlcolor=ppink]{hyperref}
\else     
  \usepackage[dvipdfmx]{}     
  \usepackage[dvipdfmx,bookmarksopen,colorlinks=true,linkcolor=bblue,citecolor=ppink,urlcolor=darkred]{hyperref}
\fi

\definecolor{red}{rgb}{1,0,0}
\definecolor{darkred}{rgb}{0.6,0,0}
\definecolor{darkgreen}{rgb}{0.992447,0.623778,0.034597}
\definecolor{ppink}{rgb}{1,0.4,0.4}
\definecolor{bblue}{rgb}{0.284602,0.317763,0.963947}

\newcommand{\footnoteref}[1]{\protected@xdef\@thefnmark{\ref{#1}}\@footnotemark}

\makeatother

\begin{document}

\preprint{IPMU17-0142}

\title{Domain wall and isocurvature perturbation problems \\in a supersymmetric
axion model}

\author{Masahiro Kawasaki}

\affiliation{ICRR, University of Tokyo, Kashiwa, 277-8582, Japan}

\affiliation{Kavli IPMU (WPI), UTIAS, University of Tokyo, Kashiwa, 277-8583,
Japan}

\author{Eisuke Sonomoto}

\affiliation{ICRR, University of Tokyo, Kashiwa, 277-8582, Japan}

\affiliation{Kavli IPMU (WPI), UTIAS, University of Tokyo, Kashiwa, 277-8583,
Japan}
\begin{abstract}
\noindent The axion causes two serious cosmological problems, domain
wall and isocurvature perturbation problems. Linde pointed that the
isocurvature perturbations are suppressed when the Peccei-Quinn (PQ)
scalar field takes a large value $\sim M_{\text{pl}}$ (Planck scale) during inflation.
In this case, however, the PQ field with large amplitude starts to oscillate
after inflation and large fluctuations of the PQ field are produced through parametric
resonance, which leads to the formation of domain walls. We consider a supersymmetric
axion model and examine whether domain walls are formed by using lattice
simulation. It is found that the domain wall problem does not appear
in the SUSY axion model when the initial value of the PQ field is
less than $10^{3}\times v$ where $v$ is the PQ symmetry breaking
scale. 
\end{abstract}
\maketitle


\section{Introduction}

\label{sec:introduction} 

The axion~\cite{Weinberg:1977ma,Wilczek:1977pj}
is a scalar particle which results from the Peccei-Quinn (PQ) mechanism~\cite{Peccei:1977ur,Peccei:1977hh},
the most attractive solution to the strong CP problem~\cite{tHooft:1976rip}.
In the PQ mechanism we impose new $U(1)$ symmetry on the standard
model, called $U(1)_{\text{PQ}}$ symmetry. The axion is a Nambu-Goldstone
boson generated by the spontanous symmetry breaking of the $U(1)_{{\rm PQ}}$
symmetry. The axion is also an attractive candidate for the cold dark
matter (CDM)~\cite{Preskill:1982cy,Abbott:1982af,Dine:1982ah} because
its coherent oscillation behaves like nonrelativistic matter. Thus,
the axion is very fascinating in that it solves two important
problems in particle physics and cosmology simultaneously.

However, the axion causes two serious cosmological problems, depending
on the epoch of the PQ symmetry breaking. If the symmetry breaking
occurs after inflation, the domain wall problem arises~\cite{Sikivie:1982qv}.
When the univerese cools down to the QCD scale, the axion potential
is lifted up by the QCD instanton effect and the axion acquires 
mass, typically $m_{a}\sim10^{-4}$ eV. After the axion potential is
formed, the axion field starts to roll down the potential toward its
minima. At that time, domain walls are formed because the axion potential has $N$ discrete
minima ($N\in Z$ is the QCD anomaly factor depending on the axion
model) and the initial values of the axion field are spatially random. For the axion model with $N=1$, the formed
domain walls are disk-like objects whose boundaries are axionic strings
produced by the $U(1)_{{\rm PQ}}$ symmetry breaking, and they soon
collapse because of their surface tension~\cite{Vilenkin:1982ks}.
This leads to the axion overproduction unless $F_{a}\lesssim10^{10}$~GeV~\cite{Hiramatsu:2012gg}.
For $N\geq2$, stable domain walls are formed and their energy density
soon dominates the universe~\cite{Ryden:1989vj}.

On the other hand, if the PQ symmetry breaking occurs before or during
inflation, the isocurvature perturbation problem arises~\cite{Axenides:1983hj,Seckel:1985tj,Linde:1985yf,Linde:1990yj,Turner:1990uz,Lyth:1991ub}.
Because a tiny region where the axion field is uniform expands into the region larger than the present Hubble horizon by inflation, the axion field takes
almost the same value in the whole observable universe. Thus, no domain walls
are produced. However, during inflation the axion field $a$ acquires
quantum fluctuations represented by $\delta a=H_{{\rm inf}}/(2\pi)$
with $H_{{\rm inf}}$ being the Hubble parameter during inflation. Here we
introduce the phase of the PQ field $\theta$ and misalignment angle
$\theta_{a}$. They are defined as  $\theta \equiv a/v$ and $\theta_{a} \equiv \theta\,N=a/F_{a}$ where $v$ is the PQ breaking scale and $F_{a}(=v/N)$ is the axion
decay constant. Then the misalignment angle fluctuation is written
as 
\begin{equation}
\frac{\delta\theta_{a}}{\theta_{a}}=\frac{\delta a}{F_{a}\theta_{a}}=\frac{H_{{\rm inf}}}{2\pi F_{a}\theta_{a}}.
\end{equation}
In high scale inflation like chaotic inflation, $H_{{\rm inf}}\gtrsim10^{13}$~GeV, so $\delta\theta_{a}/\theta_{a}\sim O(1)$ for $F_{a}\sim10^{12}$~GeV.
After the axion obtains mass, the axion fluctuations lead to the
isocurvature density perturbations with the amplitude $\sim\delta\theta_{a}/\theta_{a}$.
Because the isocurvature perturbations are stringently constrained
by the cosmic microwave background (CMB) observations~\cite{Ade:2015lrj},
the axion causes a serious cosmological problem unless
the Hubble parameter during inflation is small.

Linde pointed out that these two problems are
solved simultaneously when the PQ field $\Phi$ takes a large expectation
value $|\Phi_{i}|\sim M_{\text{pl}}$ during inflation~\cite{Linde:1991km}. In this case
the axion misalignment fluctuations are suppressed as 
\begin{equation}
\delta\theta_{a}=\frac{\delta a}{|\Phi_{i}|/N}=\frac{H_{\text{inf}}}{2\pi|\Phi_{i}|/N}\sim\frac{H_{\text{inf}}}{2\pi M_{\text{pl}}},
\end{equation}
which is small even for high scale inflation. Therefore, the isocurvature
perturbation problem does not appear.

However, after inflation the PQ field starts to oscillate and the
fluctuations of the PQ field grow through parametric resonance~\cite{Kofman:1995fi,Kofman:1997yn,Shtanov:1994ce}.
If these fluctuations are sufficiently large, $U(1)_{\text{PQ}}$
symmetry is non-thermally restored~\cite{Tkachev:1998dc,Kasuya:1998td,Kasuya:1999hy,Kawasaki:2013iha} and the domain walls are produced after QCD phase transition.
Most recently, Ref.~\cite{Kawasaki:2013iha} examined whether the PQ symmetry is
restored and domain walls are formed by using lattice simulation.
It was found that the initial value of the PQ field $\Phi$ should
satisfy 
\begin{equation}
|\Phi|_{i}\lesssim F_{a}\times10^{4}.
\end{equation}
to avoid the domain wall formation.

Linde's idea also works in the supersymmetric (SUSY) axion model where
the flat direction of the PQ scalar potential plays the role of the
PQ field. In Ref.~\cite{Kasuya:1996ns} it was shown that the scalar field corresponding
to the flat direction starts to oscillate by SUSY breaking mass
terms and it causes parametric resonance after inflation, which may
lead to the domain wall formation. However, the resonance effect was
only studied in linear analysis and it is uncertain whether domain
walls are actually formed. Therefore, in this paper we examine the
formation of domain walls in the SUSY axion model by using lattice
simulation. 
We perform the lattice simulations for the case that the initial value
of the PQ field is larger than $v$ at most by a factor $10^{3}$
and find that the axion fluctuations generated by parametric resonance
are not large enough to produce domain walls.

Sec.~\ref{sec:Supersymmetric-axion-model} introduces the SUSY axion
model that we use in the lattice simulations. Sec.~\ref{sec:Simulation-setup}
derives equations of motion and some other equations that are necessary
for the simulations and explain the lattice condition. Sec. \ref{sec:Result-of-the}
describes the result of the numerical simulations. Finally Sec.~\ref{sec:Conclusion-and-discussion}
summarizes our result. 


\section{Supersymmetric axion model \label{sec:Supersymmetric-axion-model}}


We adopt a SUSY axion model with superpotential represented by 
\begin{equation}
W=h(\Psi_{+}\Psi_{-}-v^{2})\Psi_{0}.
\end{equation}
where $\Psi_{+},$ $\Psi_{-},$ $\Psi_{0}$ are chiral superfields
with PQ charge $+1,$ $-1,$ $0$, respectively, and the coupling
constant $h$ is assumed to be $O(1)$. Then the scalar potential
is 
\begin{align}
V_{{\rm SUSY}} & =\left|\frac{\delta W}{\delta\Psi_{i}}\right|_{\Psi_{i}\rightarrow\Phi_{i}}^{2},\nonumber \\
 & =h^{2}|\Phi_{+}\Phi_{-}-v^{2}|^{2}+h^{2}(|\Phi_{+}|^{2}+|\Phi_{-}|^{2})|\Phi_{0}|^{2}.
\end{align}
where $\Phi_{+},$ $\Phi_{-},$ $\Phi_{0}$ are scalar components
of $\Psi_{+},$ $\Psi_{-},$ $\Psi_{0}$, respectively. In this potential,
the field satisfying 
\begin{equation}
\Phi_{+}\Phi_{-}=v^{2},~~~~\Phi_{0}=0,\label{eq:flat_direction}
\end{equation}
is called flat direction. 
We also introduce soft SUSY breaking mass terms to the potential 
\begin{equation}
V=V_{{\rm SUSY}}+m_{+}^{2}|\Phi_{+}|^{2}+m_{-}^{2}|\Phi_{-}|^{2}+m_{0}^{2}|\Phi_{0}|^{2}.
\end{equation}
where $m_{\pm},\ m_{0}$ are soft masses of $O(1)$ TeV. The flat
direction is lifted by the SUSY breaking mass terms
\footnote{The flat direction is also lifted up by quantum corrections. However, it is logarithmic potential as $m_+^2\left(1+\kappa \log|\Phi _+|/M  \right)|\Phi _+|^2\ (\kappa >0)$ in the SUSY version of axion modes and does not affect the dynamics of PQ fields.}
and has a minimum at 
\begin{align}
|\Phi_{+}| & =\sqrt{\frac{m_{-}}{m_{+}}}v,\\
|\Phi_{-}| & =\sqrt{\frac{m_{+}}{m_{-}}}v.
\end{align}
Generally we should also consider Hubble induced mass terms in the
scalar potential $V$~\cite{Dine:1995uk}, but in this paper we assume
such terms are suppressed by some symmetry
\footnote{The following discussion is also applicable even if we consider the Hubble induced mass terms. When Hubble induced mass terms like
\[ V =  V_\text{SUSY} + c_{+}H^2|\Phi_{+}|^2 ++ c_{-}H^2|\Phi_{-}|^2 + c_{0}H^2|\Phi_{0}|^2, \]
are introduced, parametric resonance occurs depending on the sign of the $\mathcal{O}(1)$ coefficients $c_\pm ,c_0$. When $c_{+} < 0$, $c_{-} >0$ and $c_0 >0$ during inflation, for example, the scalar fields settle down to the flat direction satisfying $|\Phi_{+}| \simeq M_p \gg |\Phi_{-}|\simeq v^2/M_p$. Thus, in this case, parametric resonance occurs after inflation.}~\cite{Gaillard:1995az}.

To solve the strong CP problem, the PQ fields must have the interaction with the quark sector. Here, we consider two models. 

The first one is the SUSY version of KSVZ axion model~\cite{Kim:1979if,Shifman:1979if}. In this model, the PQ field $\Psi_+$ interacts with heavy quark as
\begin{align}
W_{\rm KSVZ}=k\Psi _+\bar{Q}Q,
\end{align}
where $Q$ and $\bar{Q}$ are the chiral superfields transformed as fundamental and anti-fundamental representations
of SU(3)$_C$ with PQ charge $-1/2$. The domain wall number of this model depends on the kinds of heavy quarks and the minimum one is $N=1$.

The second is the SUSY version of DFSZ axion model~\cite{Dine:1981rt,Zhitnitsky:1980tq}. In this model, the PQ field $\Psi_+$ interacts with Higgs fields as
\begin{align}
W_{\rm DFSZ}= \lambda \frac{\Psi_+^2}{M_{\rm pl}} H_u H_d,
\end{align}
where $H_u$ and $H_d$ are the up- and down-type Higgs doublets.
After the PQ field takes the expectation value $v$, the coefficient of the $\mu$ term becomes
\begin{equation}
\mu \sim \lambda \frac{v^2}{M_{\rm pl}} \sim \mathcal{O}(1)\ {\rm TeV},
\end{equation}
for $\lambda \sim \mathcal{O}(1)$ and $v\sim 10^{11}$ GeV.
Thus, in this model, the PQ scale is closely related to the solution to the $\mu$ problem in MSSM~\cite{Kim:1983dt}. The domain wall number of this model is $N=6$.


\subsection{Axion in SUSY axion model}


In the SUSY axion model, the axion field is a combination of the phases of $\Phi_{+}$ and $\Phi_{-}$. To write the axion field explicitly, we decompose $\Phi_{\pm}$
as 
\begin{equation}
\Phi_{\pm}\equiv \varphi_{\pm}\exp\left(i\frac{a_{\pm}}{\varphi_{\pm}}\right).\label{eq:decompose}
\end{equation}
and define fields $a$ and $b$ as 
\begin{align}
a & =\frac{1}{(\varphi_{+}^{2}+\varphi_{-}^{2})^{1/2}}(\varphi_{+}a_{+}-\varphi_{-}a_{-}),\label{eq:axion_field}\\[0.6em]
b & =\frac{1}{(\varphi_{+}^{2}+\varphi_{-}^{2})^{1/2}}(\varphi_{-}a_{+}+\varphi_{+}a_{-}).
\end{align}
Then for $\Phi _0=0$ the potential is rewritten as 
\begin{align}
V_{{\rm SUSY}} & =\left|\varphi_{+}\varphi_{-}\exp\left[i\left(\frac{a_{+}}{\varphi_{+}}+\frac{a_{-}}{\varphi_{-}}\right)\right]-v^{2}\right|^{2},\nonumber \\[0.6em]
 & =\left|\left[\varphi_{+}\varphi_{-}\cos\frac{(\varphi_{+}^{2}+\varphi_{-}^{2})^{1/2}}{\varphi_{+}\varphi_{-}}b-v^{2}\right]+i\varphi_{+}\varphi_{-}\sin\left(\frac{a_{+}}{\varphi_{+}}+\frac{a_{-}}{\varphi_{-}}\right)\right|^{2},\nonumber \\[0.6em]
 & =v^{4}+(\varphi_{+}\varphi_{-})^{2}-2v^{2}\varphi_{+}\varphi_{-}\cos\frac{(\varphi_{+}^{2}+\varphi_{-}^{2})^{1/2}}{\varphi_{+}\varphi_{-}}b,
\end{align}
which shows that $a$ is massless and $b$ has mass $O(v)$ at the
potential minimum. Thus $a$ is identified as the axion field.


\subsection{Cosmological evolution of axion}


During inflation the scalar fields $\Phi_{+},\Phi_{-}$ and $\Phi_{0}$
roll down to the flat direction represented by Eq.~(\ref{eq:flat_direction}),
and the field value along the flat direction can be as large as the
Planck mass $M_{\text{pl}}$. Assuming $\Phi_{+}>\Phi_{-}$, $\Phi_{\pm}$
are 
\begin{align}
|\Phi_{+}|=\varphi_{+} & \simeq M_{\text{pl}},\label{eq:PQ_field_inf+}\\
|\Phi_{-}|=\varphi_{-} & \simeq\frac{v^{2}}{M_{\text{pl}}},\label{eq:PQ_field_inf-}
\end{align}
which leads to solve the isocurvature perturbation problem as seen
later. Soft mass terms are negligible in this epoch because they are
much smaller than the Hubble parameter. After inflation, the soft
SUSY breaking masses become comparable to the Hubble parameter and
$\Phi_{+}$ and $\Phi_{-}$ start to oscillate along the flat direction.
Hereafter, we call the scalar degree of freedom corresponding to the
flat direction as PQ field which is the combination of $\Phi_{+}$
and $\Phi_{-}$. For $|\Phi_{+}| \gg |\Phi_{-}|$ the PQ field is $\simeq \Phi_{+}$. During oscillation of the PQ field, the axion fluctuations
grow through parametric resonance.
However, unlike in the non-SUSY axion model, in the SUSY axion model
the PQ symmetry is not restored non-thermally through parametric resonance.
To restore the $U(1)_{PQ}$ symmetry in this model, the fluctuations
of the sclar fields orthogonal to the flat direction, which we call
the orthogonal direction, must be $O(v)$. Yet, because the mass of
the orthogonal direction ($\sim v$) is much larger than the oscillation
scale ($\sim m_{\pm}$), its fluctuations do not grow through parametric
resonance. Thus, the $U(1)_{\text{PQ}}$ symmetry is not restored
in this SUSY axion model.

On the other hand, fluctuations along the flat direction, i.e., PQ field
including axion, can be large through parametric resonance as pointed
out in Ref.~\cite{Kasuya:1996ns}. If the axion fluctuations become
as large as $\delta\theta_{a}\sim O(1)$, domain walls may be formed
at the QCD epoch and we may confront the domain wall problem. Because
the $U(1)_{\text{PQ}}$ is not restored, the formed domain walls have
no boundaries, so they are stable even for $N=1$.


\subsection{Suppression of isocurvature perturbations}

\begin{figure}[H]
\centering\includegraphics[scale=0.5]{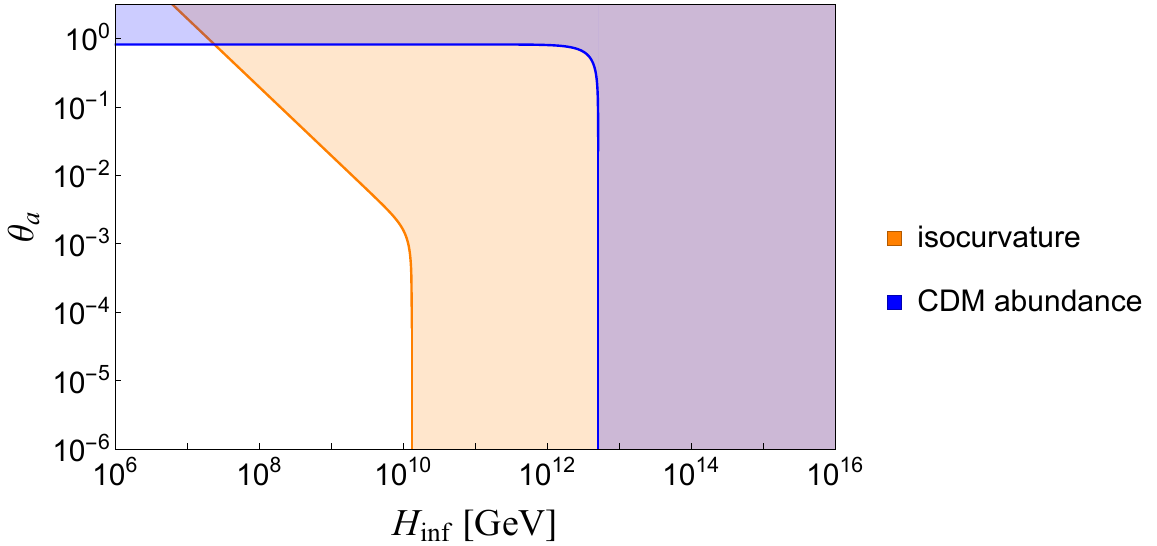}
\caption{Constraints on $H_{{\rm inf}}$ and $\theta_{a}$ when we assume $F_{a}=10^{12}$
GeV. The orange and blue regions are excluded by the observation of
the isocurvature perturbations and cold dark matter abundance respectively.}
\label{no_Linde}
\end{figure}

Let us estimate the isocurvature perturbations in the SUSY axion model.
During inflation $\Phi_{+}$ and $\Phi_{-}$ have field values represented
by Eqs.~(\ref{eq:PQ_field_inf+}) and (\ref{eq:PQ_field_inf-}).
For $|\Phi_{+}|\gg|\Phi_{-}|$, the axion field is almost identical
to the phase $\theta_{+}$ of $\Phi_{+}$ {[}see Eq.~(\ref{eq:axion_field}){]},
\begin{equation}
a\simeq a_{+}\equiv\theta_{+}\varphi_{+}.
\end{equation}
The axion field $a_{+}$ obtains fluctuations of $H_{\text{inf}}/(2\pi)$
during inflation and it leads to the fluctuations of the phase angle
$\delta\theta_{+}$ given by 
\begin{equation}
\delta\theta_{+}(\simeq-\delta\theta_{-})=\frac{H_{\text{inf}}}{2\pi\varphi_{+,i}},
\end{equation}
where $\varphi_{+,i}$ is the value of $\varphi_{+}$ during inflation.
Taking into account that the phase angle is preserved until the start
of the axion oscillation, after the scalar fields $\varphi_{+}$ and $\varphi_{-}$ settle at the minimum
of the potential, the fluctuation of the axion misalignment angle
is obtained as 
\begin{equation}
\delta\theta_{a}=\frac{N}{\varphi_{+}^{2}+\varphi_{-}^{2}}(\varphi_{+}^{2}\delta\theta_{+}-\varphi_{-}^{2}\delta\theta_{-})\simeq N\delta\theta_{+}.
\end{equation}
The CDM isocurvature perturbations $S_{a}$ due to the axion field is written
as 

\begin{equation}
S_{a}=\frac{\Omega_{a}}{\Omega_{{\rm CDM}}}\frac{2\theta_{a}\delta\theta_{a}+\delta\theta_{a}^{2}-\langle\delta a^{2}\rangle}{\theta_{a}^{2}+\langle\delta\theta_{a}^{2}\rangle}.
\end{equation}
Here, we assume that $\delta\theta_{a}$ obeys Gaussian distribution, and
 $\Omega_{a}$ and $\Omega_{{\rm CDM}}$ are the density parameters
of the axion and cold dark matter. They are 
\begin{align}
\Omega_{a}h^{2} & =0.18\times\left(\theta_{a}^{2}+\langle\delta\theta_{a}^{2}\rangle\right)\left(\frac{F_{a}}{10^{12}\text{GeV}}\right)^{1.19},\\
\Omega_{{\rm CDM}}h^{2} & =0.12.\label{eq:axion_density}
\end{align}
 From the CMB observations \cite{Ade:2015lrj} the power spectrum
of the isocurvature perturbations is stringently constrained as 

\begin{equation}
\beta_{{\rm iso}}\equiv\frac{\mathcal{P}_{{\rm iso}}(k_{0})}{\mathcal{P}_{{\rm iso}}(k_{0})+\mathcal{P}_{{\rm ad}}(k_{0})}<0.038
\end{equation}
where $\mathcal{P}_{{\rm iso}}(k)$ and $\mathcal{P}_{{\rm ad}}(k)$
are amplitudes of isocurvature and adiabatic fluctuations,
and $k_{0}=0.05\ {\rm Mpc}^{-1}$ is the pivot scale. Using $\mathcal{P}_{{\rm ad}}(k_{0})=2.20\times10^{-9}$
\cite{Ade:2015lrj}, $\mathcal{P}_{{\rm iso}}$ must satisfy

\begin{equation}
\mathcal{P}{}_{\text{iso}}\equiv\langle|S_{a}|^{2}\rangle=\left(\frac{\Omega_{a}}{\Omega_{{\rm CDM}}}\right)^{2}\frac{2\langle\delta\theta_{a}^{2}\rangle(2\theta_{a}^{2}+\langle\delta\theta_{a}^{2}\rangle)}{(\theta_{a}^{2}+\langle\delta\theta_{a}^{2}\rangle)^{2}}<8.7\times10^{-11}.
\end{equation}
Then the Hubble parameter during inflation should satisfy 
\begin{equation}
\left[2\theta_{a}^{2}+\left(\frac{NH_{{\rm inf}}}{2\pi\varphi_{+,i}}\right)^{2}\right]\left(\frac{NH_{{\rm inf}}}{2\pi\varphi_{+,i}}\right)^{2}\left(\frac{F_{a}}{10^{12}\ {\rm GeV}}\right)^{2.38}\lesssim1.9\times10^{-11}.\label{eq:iso}
\end{equation}
For the case of $\varphi_{+,i}\simeq v=F_{a}N$ (which represents the standard case without suppression of the axion fluctuation),  
\begin{equation}
\left[2\theta_{a}^{2}+\left(\frac{H_{{\rm inf}}}{2\pi F_{a}}\right)^{2}\right]\left(\frac{H_{{\rm inf}}}{2\pi F_{a}}\right)^{2}\left(\frac{F_{a}}{10^{12}\ {\rm GeV}}\right)^{2.38}\lesssim1.9\times10^{-11}.
\end{equation}
In this case the constraints on $H_{{\rm inf}}$ and $\theta_{a}$
for $F_{a}=10^{12}$ GeV are shown in Fig.~\ref{no_Linde}. 
From the figure it is found the constraint from the isocurvature perturbations is very stringent and Hubble parameter during inflation should be smaller than $10^{10}$~GeV. Moreover, if the axion is dark matter, $H_\text{inf} \lesssim 10^7$~GeV. 

\begin{figure}[H]
\centering%
\begin{minipage}[t]{0.45\columnwidth}%
\includegraphics[scale=0.42]{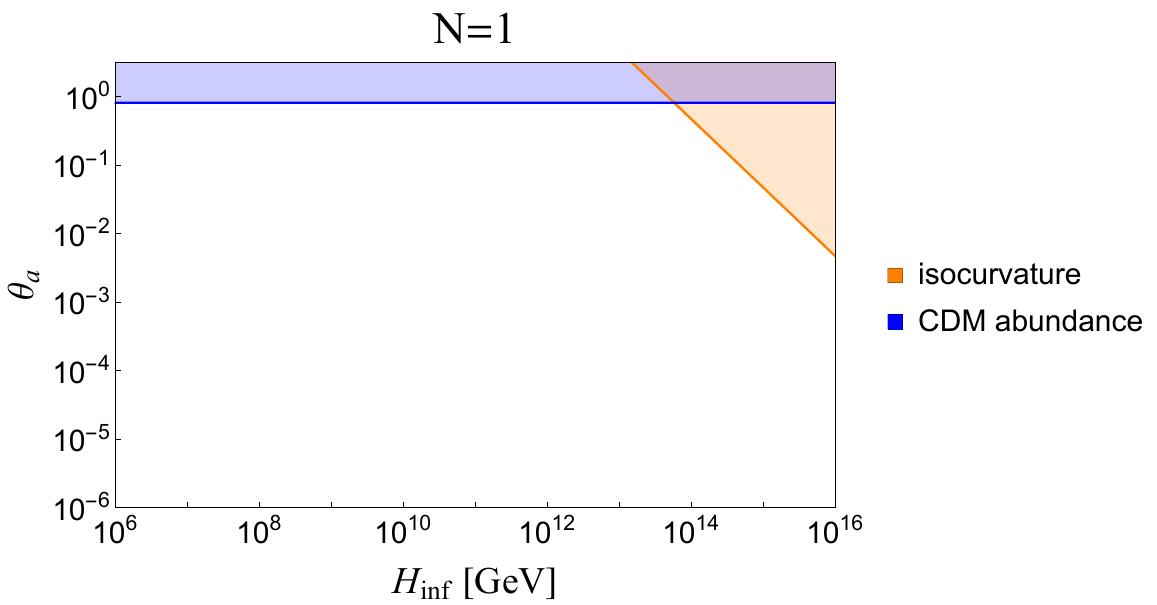}%
\end{minipage}%
\begin{minipage}[t]{0.45\columnwidth}%
\includegraphics[scale=0.42]{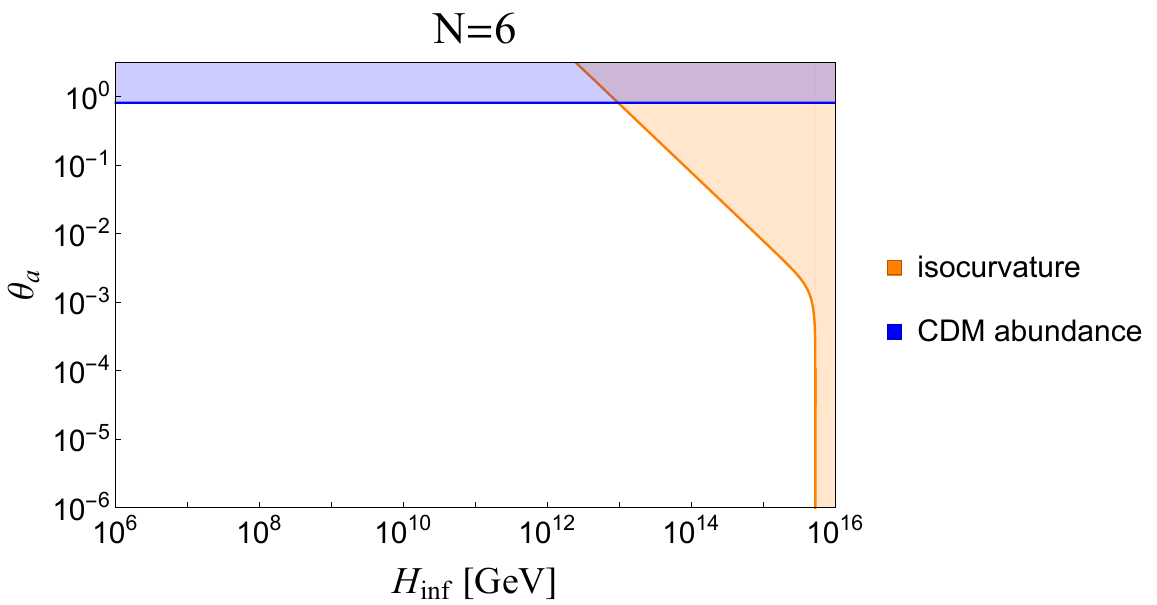}%
\end{minipage}\caption{The same constraints as Fig. \ref{no_Linde} but with the large value
$M_{{\rm pl}}$ of the PQ field during inflation. The left figure
shows the $N=1$ case and the right figure shows the $N=6$ case,
e.g. the DFSZ model.}
\label{Linde}
\end{figure}

However, if the flat direction has a large field value ($\gg F_{a}$)
the isocurvature perturbations are suppressed. In fact, if we take
$\varphi_{+,i}=M_{\text{pl}}\simeq2.43\times10^{18}$ GeV, the constraint
on $H_{\text{int}}$ is relaxed as shown in Fig.~\ref{Linde}. Yet
the domain wall formation by large fluctuations of the misalignment
angle due to parametric resonance might
exclude some of the allowed region in Fig.~\ref{Linde}. 

\section{Simulation setup\label{sec:Simulation-setup}}


As seen in the previous section, in the SUSY axion model the isocurvature perturbations are suppressed
if the PQ field has a large field value during inflation. However,
in this case the PQ field starts to oscillate after inflation and
the axion fluctuations grow through parametric resonance. If the fluctuations
become sufficiently large, the axion field takes random values in space,
which results in the domain wall formation. To examine whether
domain walls are formed we perform numerical lattice simulations.

For simplicity, we make the following assumptions through the simulations.
\begin{equation}
\Phi_{0}=0,~~~~m\equiv m_{+}=m_{-},~~~~~h=1.
\end{equation}
Then the scalar potential $V$ is written as 
\begin{equation}
V=|\Phi_{+}\Phi_{-}-v^{2}|^{2}+m^{2}(|\Phi_{+}|^{2}+|\Phi_{-}|^{2}).
\end{equation}


\subsection{Equations of motion}


We decompose complex scalar fields into their real parts and complex
parts, 
\begin{align}
\Phi_{+} & =\frac{1}{\sqrt{2}}(\phi_{0}+i\phi_{1}),\label{eq:phi+}\\
\Phi_{-} & =\frac{1}{\sqrt{2}}(\phi_{2}+i\phi_{3}).\label{eq:phi-}
\end{align}
We solve the evolution of the scalar fields taking the cosmic expansion
into account. Using the FRW metric $g_{\mu\nu}={\rm diag}(1,\ -R,\ -R,\ -R)$
where $R$ is the scale factor, equations of motion are written as
\begin{equation}
\ddot{\phi}_{i}+3H\dot{\phi}_{i}-\frac{1}{R^{2}}\Delta\phi_{i}+\frac{\partial V(\phi)}{\partial\phi_{i}}=0.\ \ (i=0,1,2,3).
\end{equation}
Furthermore, let us rescale the variables as 
\begin{align}
d\tau & =m\frac{dt}{R},\\
d\tilde{x} & =mdx,\\
\phi_{i} & =\frac{m}{R}\varphi_{i}.
\end{align}
Then equations of motion are rewritten as 
\begin{equation}
\varphi_{i}''-\tilde{\Delta}\varphi_{i}-\frac{R''}{R}\varphi_{i}+\frac{\partial V(\varphi)}{\partial\varphi_{i}}=0.
\end{equation}
where 
\begin{align}
\frac{\partial V(\varphi)}{\partial\varphi_{0}} & =\frac{1}{2}\left[\left(\varphi_{0}\varphi_{2}-\varphi_{1}\varphi_{3}-2\left(\frac{v}{m}R\right)^{2}\right)\varphi_{2}+(\varphi_{0}\varphi_{3}+\varphi_{1}\varphi_{2})\varphi_{3}\right]+a^{2}\varphi_{0},\\[0.6em]
\frac{\partial V(\varphi)}{\partial\varphi_{1}} & =\frac{1}{2}\left[-\left(\varphi_{0}\varphi_{2}-\varphi_{1}\varphi_{3}-2\left(\frac{v}{m}R\right)^{2}\right)\varphi_{3}+(\varphi_{0}\varphi_{3}+\varphi_{1}\varphi_{2})\varphi_{2}\right]+a^{2}\varphi_{1},\\[0.6em]
\frac{\partial V(\varphi)}{\partial\varphi_{2}} & =\frac{1}{2}\left[\left(\varphi_{0}\varphi_{2}-\varphi_{1}\varphi_{3}-2\left(\frac{v}{m}R\right)^{2}\right)\varphi_{0}+(\varphi_{0}\varphi_{3}+\varphi_{1}\varphi_{2})\varphi_{1}\right]+a^{2}\varphi_{2},\label{eq:varphi2}\\[0.6em]
\frac{\partial V(\varphi)}{\partial\varphi_{3}} & =\frac{1}{2}\left[-\left(\varphi_{0}\varphi_{2}-\varphi_{1}\varphi_{3}-2\left(\frac{v}{m}R\right)^{2}\right)\varphi_{1}+(\varphi_{0}\varphi_{3}+\varphi_{1}\varphi_{2})\varphi_{0}\right]+a^{2}\varphi_{3}.
\end{align}
The prime denotes the derivative with respect to $\tau$. We solve
these equations numerically in the lattice simulation.


\subsection{Relations between PQ fields and axion}


Next, we describe the relations between $\varphi_{0},\ \varphi_{1},\ \varphi_{2},\ \varphi_{3}$
and $\varphi_{\pm},\ a_{\pm}$. 
They are written as 
\begin{align}
\varphi_{+}^{2} & =\left(\frac{1}{\sqrt{2}}\frac{m}{R}\right)^{2}(\varphi_{0}^{2}+\varphi_{1}^{2}),\\
\varphi_{-}^{2} & =\left(\frac{1}{\sqrt{2}}\frac{m}{R}\right)^{2}(\varphi_{2}^{2}+\varphi_{3}^{2}).
\end{align}
from Eqs.~(\ref{eq:decompose}), (\ref{eq:phi+}), and (\ref{eq:phi-}).
Then from Eq.~(\ref{eq:axion_field}),
\begin{equation}
a=\varphi_{\pm}{\rm tan^{-1}\frac{\varphi_{1}}{\varphi_{0}}=\varphi_{\pm}{\rm tan^{-1}\frac{\varphi_{3}}{\varphi_{2}}}}.
\end{equation}
Thus the phase angle of the axion field is 
\begin{align}
\theta_{a} & =\frac{a}{(\varphi_{+}^{2}+\varphi_{-}^{2})^{1/2}},\\
 & =\left(\frac{1}{\sqrt{2}}\frac{m}{R}\right)^{-1}\frac{1}{\varphi_{0}^{2}+\varphi_{1}^{2}+\varphi_{2}^{2}+\varphi_{3}^{2}}\left[(\varphi_{0}^{2}+\varphi_{1}^{2})\tan^{-1}\frac{\varphi_{1}}{\varphi_{0}}-(\varphi_{2}^{2}+\varphi_{3}^{2})\tan^{-1}\frac{\varphi_{3}}{\varphi_{2}}\right].
\end{align}

\subsection{Simulation condition}

In reality, the typical mass scales of $m$ and $v$ are about $10^{3}$~GeV
and $10^{12}$~GeV, respectively. However, in numerical simulation
it is impossible to study the dynamical system with such hugely different
mass scales at the same time. Therefore, we set $v/m$ being as small as to
satisfy the condition that the scalar fields are confined in the flat
direction. Suppose that $A$ is the initial 
amplitude of the homogeneous part of $\Phi_{+}$. Then, because the energy density of the orthgonal direction and
the PQ field is about $v^{4}$ and $m^{2}(Av)^{2}$, $A$ should
satisfy

\begin{align}
A & <\frac{v}{m}.
\end{align}

As for the initial conditions for the scalar fields we take 
\begin{align}
\langle{\rm Re}\Phi_{+}\rangle_{i} & =A\times v,\\
\langle{\rm Re}\Phi_{-}\rangle_{i} & =\frac{1}{A}\times v,\\
\langle{\rm Im}\Phi_{+}\rangle_{i} & =\langle{\rm Im}\Phi_{-}\rangle_{i}=0.
\end{align}
where $i$ denotes the initial values and we changed $A$ in the lattice
simulations. We initially assign tiny fluctuations, which are random
numbers in $[-10^{-6}\ ,10^{6}]$, to the scalar fields at each
lattice point. We assume that the universe is matter dominated and
set the initial time $\tau_{i}=1$ and the initial scale factor $R(\tau_{i})=1$,
which means $R(\tau)=\tau^{2}$.
To study the precise evolution of the PQ field after inflation,
we have performed the simulation in 2 dimensions with periodic boundary
conditions. For the test of the code we calculate the total energy density of the system
without cosmic expansion, which is conserved with accuracy less than 0.2\% for $A=100$ and $v/m=500$.


\section{Result of the lattice simulation \label{sec:Result-of-the}}


\begin{table}[H]
\centering 
\caption{Simulation parameters. $L$ and $N$ are the box size and the number
of grid points. $\Delta\tau$ is the time step of the simulation.
The box size $L$ is normalized by the oscillation mass scale $m$. }
\begin{tabular}{crrrrr}
\hline 
 & ~~~~~~~$A$~  & ~~~~~~$v/m$  & ~~~~~~$mL$  & ~~~~~~~~~$N$~~  & ~~~~~~~~~~~$\Delta\tau$~~ \tabularnewline
\hline 
\hline 
(i)  & $100$  & $50$  & $10$  & $512^{2}$  & $2\times10^{-4}$\tabularnewline
\hline 
(ii)  & $200$  & $100$  & $10$  & $1024^{2}$  & $7\times10^{-5}$\tabularnewline
\hline 
(iii)  & $500$  & $200$  & $10$  & $2200^{2}$  & $1\times10^{-5}$\tabularnewline
\hline 
(iv)  & $1000$  & $500$  & 10  & $2048^{2}$  & $2\times10^{-6}$\tabularnewline
\hline 
\end{tabular}

\label{Ta:parameter} 
\end{table}


We perform the simulations changing $A$ and the lattice parameters
as shown in Table~\ref{Ta:parameter}. We confirm that the behavior of the system does not change with the different simulation box size $mL$. In this paper, the maximum
value of $A$ is $1000$ coming from the limit of the available machine power. 

In numerical simulation, the time step must be smaller than any oscillation time 
scales of the equations of motion. For the equations
of motion of $\varphi_{2}$ given by

\begin{equation}
\varphi_{2}''-\tilde{\Delta}\varphi_{2}-\frac{R''}{R}\varphi_{2}+\frac{\partial V(\varphi)}{\partial\varphi_{2}}=0,
\end{equation}
the most important part in the initial configuration 
is the time derivative of $\varphi_{2}$ and its potential. Then the
equation is effectively written as

\begin{align}
\varphi_{2}'' & \simeq-\frac{\partial V(\varphi)}{\partial\varphi_{2}},\\
 & =-\frac{1}{2}\left[\left(\varphi_{0}\varphi_{2}-\varphi_{1}\varphi_{3}-2\left(\frac{v}{m}R\right)^{2}\right)\varphi_{0}+(\varphi_{0}\varphi_{3}+\varphi_{1}\varphi_{2})\varphi_{1}\right]-a^{2}\varphi_{2}, \\
 & \simeq-\frac{1}{2}\left(\varphi_{0}^{2}\varphi_{2}-2\left(\frac{v}{m}R\right)^{2}\varphi_{0}\right)-a^{2}\varphi_{2}.\label{eq:simvarphi2}
\end{align}
and the largest oscillation time scale of this equation is about $|\varphi_{0}|=A\times v/m$,
the first term of Eq.~(\ref{eq:simvarphi2}). Thus if we perform
the simulations with the larger value of $A$ , we have to take smaller
value of $\Delta\tau\sim(A\times v/m)^{-1}$, which requires much
longer CPU time and hence sets the maximum value of $A$.

Because the epoch of the PQ field oscillation is much earlier than
the QCD phase transition, the axion is massless and the axion field
tends to be homogeneous inside the horizon due to the gradient term
of the equation of motion. Thus, for domain walls to be formed the
field value of the axion must be different in each Hubble patch after
parametric resonance. Because the box size $L$ of our simulations
is smaller than the horizon size $H^{-1}\sim\tau^{3}$ at the end
of the simulations, we cannot directly see that the axion field takes
different values at different Hubble patches. Thus, to examine whether
the domain wall problem arises, we investigate the spatial distribution
of the field value of the axion after parametric resonance in the
simulation box. If the axion field takes completely random value through
parametric resonance, the axion angle $\theta_{a}$ has a flat distribution
in $[-\pi,\pi]$ just after the resonance. In the subsequent evolution
the gradient term aligns the angle, which leads to the angle distribution
with a peak at some value. Once the angle becomes sufficiently random,
the peak angle is also random. In other words, if the axion fluctuations
do not grow enough, the angle distribution has a peak at the initial
value, $\langle\theta_{a}\rangle_{i}=0$.

\begin{figure}[H]
\centering \includegraphics[width=8cm]{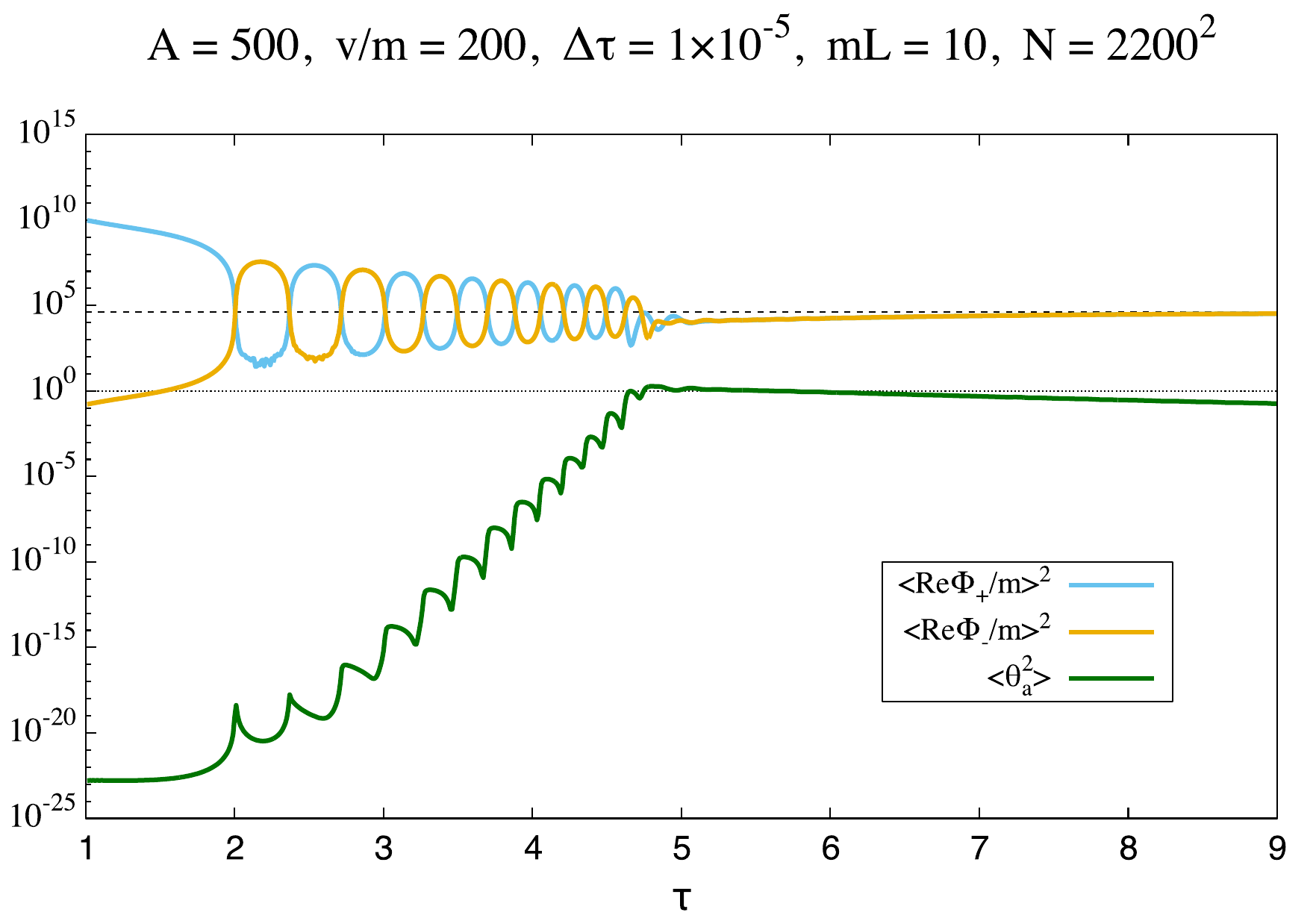} \includegraphics[width=8cm]{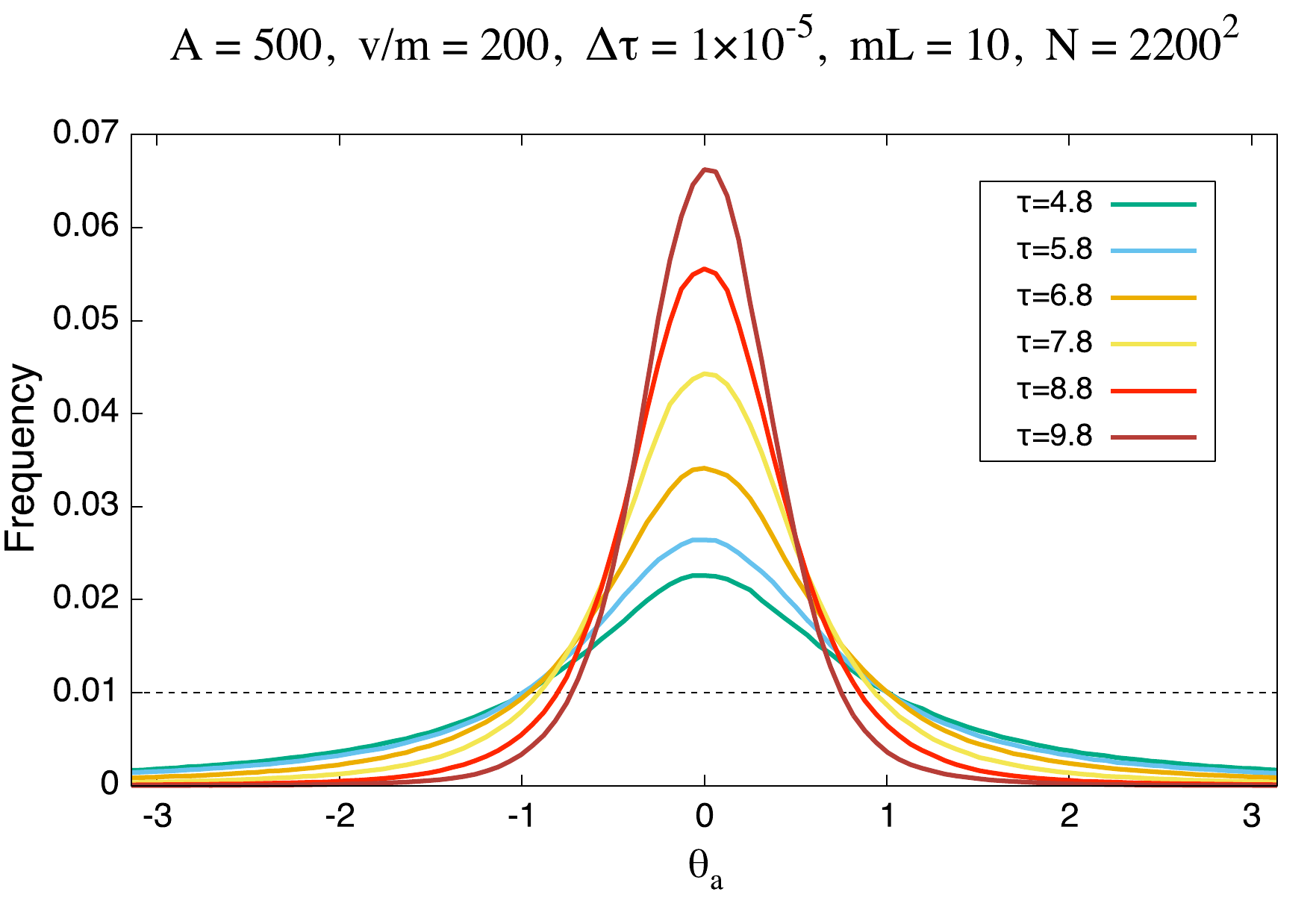}
\caption{
Left: the time evolution of the average of real parts of PQ fields,
${\rm \langle Re}\Phi_{\pm}\rangle$, and the variance of the axion
field $\langle\theta_{a}^{2}\rangle$ for the case of (iii) in the
Table~\ref{Ta:parameter}. The broken line shows $v/m$ and the dotted
line shows $10^{0}$. Right: the histogram of the spatial distribution
of the axion angle $\theta_{a}$. The distribution $f$ is normalized
as $\int_{-\pi}^{\pi}fd\theta_{a}=2\pi/100$. The dotted line shows the case where the values of $\theta _a$ are completely random. }
\label{fig:Fa_200} 
\end{figure}


\begin{figure}[H]
\centering \includegraphics[width=8cm]{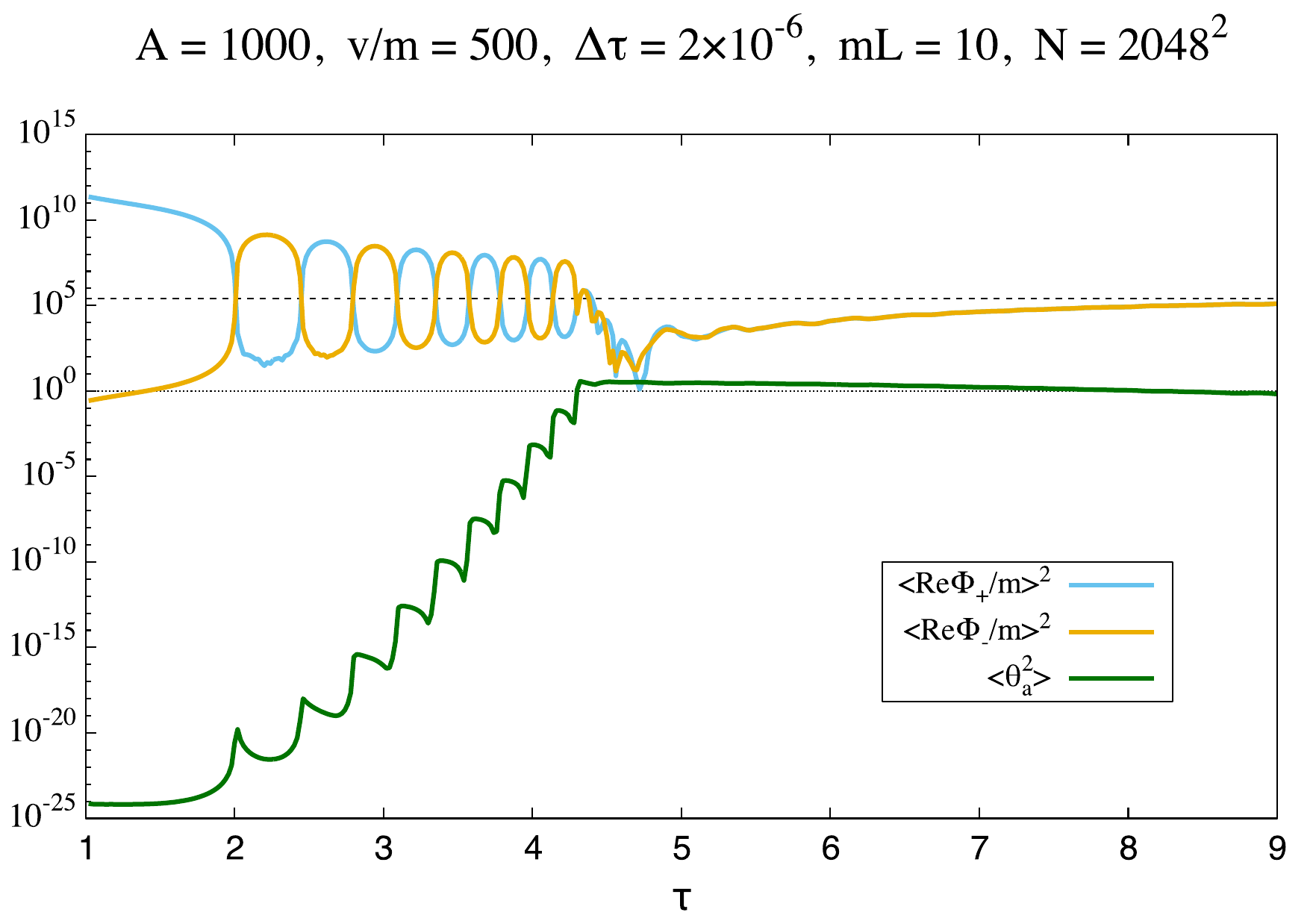} \includegraphics[width=8cm]{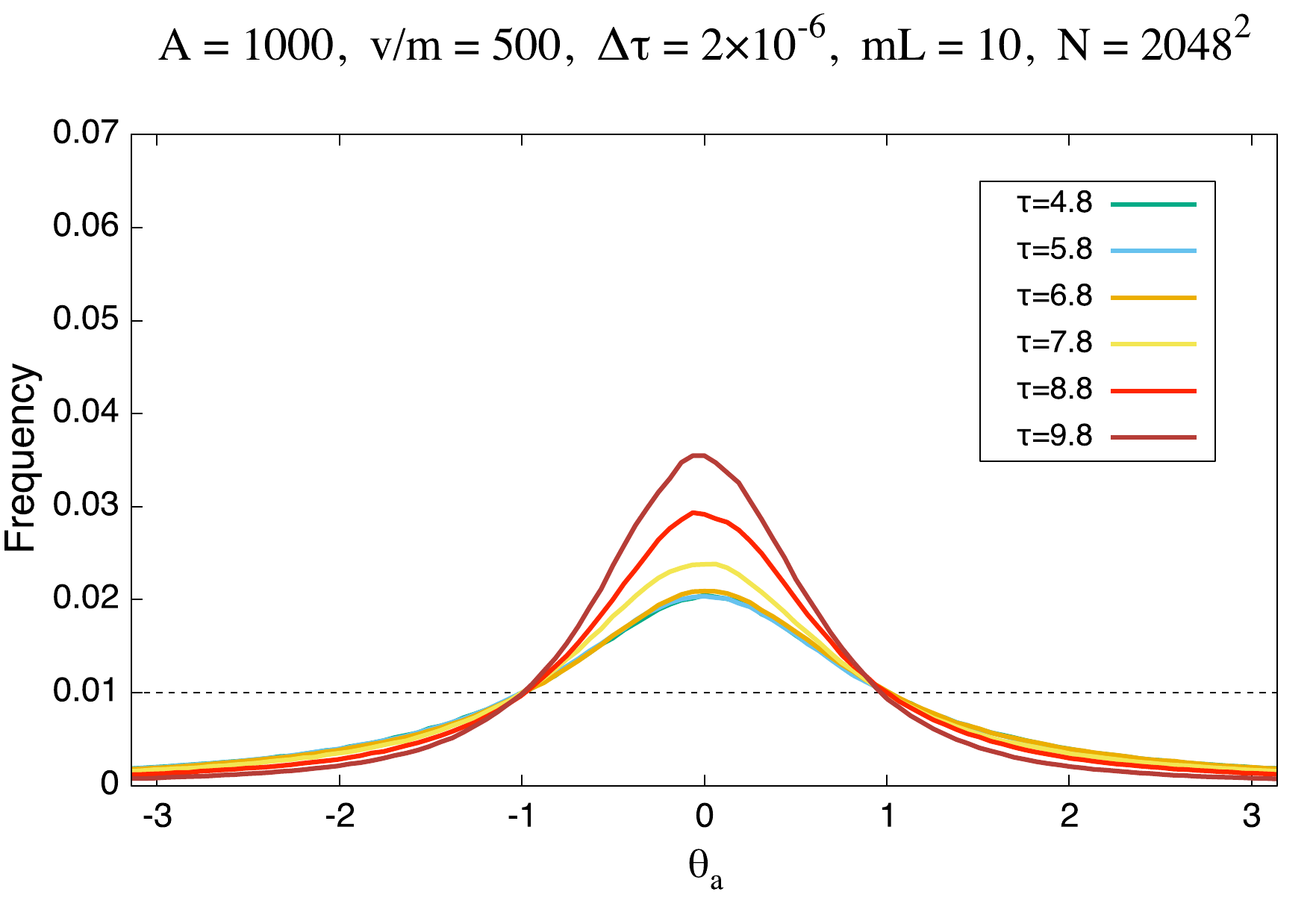}
\caption{
The same as Fig.~\ref{fig:Fa_200} but for the case of (iv) in the
Table~\ref{Ta:parameter}. }
\label{fig:Fa_500} 
\end{figure}


The result of our simulations is shown in Figs.~\ref{fig:Fa_200}
and \ref{fig:Fa_500} for $A=500$ and $A=1000$. The left panels
of the figures show the average of the $\Phi_{\pm}$ fields, $\langle{\rm Re}\Phi_{\pm}\rangle$,
and the variance of the axion angle, $\langle\theta_{a}^{2}\rangle$.
We find that the axion fluctuations become large through parametric
resonance during the $\Phi_{\pm}$ oscillating. Then
at $\tau\simeq4.8$ the angle fluctuations reach $O(1)$, that means
the angle is widely distributed between $-\pi$ and $\pi$. However,
afterward the angle is aligned by the gradient term and its fluctuation
gradually decrease. This situation is shown in the right panels of
the figures where the time evolutions of the angle distributions are
shown. In these figures, we divided $-\pi$ to $\pi$ equally into
$100$ bins and plot the frequency of the value of $\theta_{a}$ from $\tau\simeq4.8$
every $\Delta \tau =1$, which corresponds to the point of maximum fluctuations
of the axion field. We find that the values of the axion angle come
back to the initial one, $\langle\theta_{a}\rangle_{i}=0$. Therefore,
domain wall problem does not arise at least for $A\leq 1000$.

\begin{figure}[H]
\centering\includegraphics[scale=0.5]{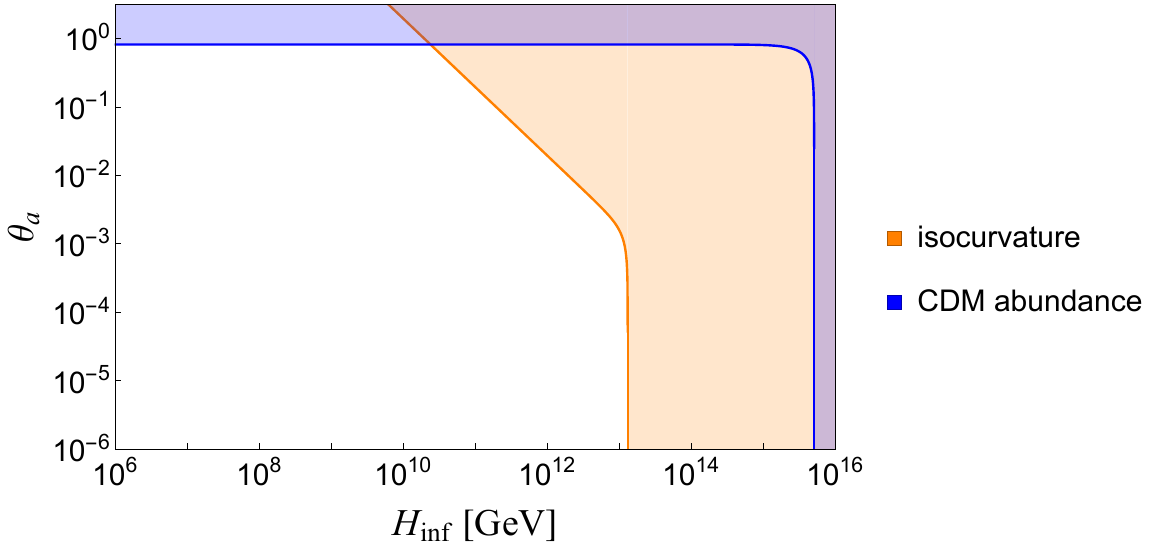}
\caption{Constraints on $H_{{\rm inf}}$ and $\theta_{a}$ when we use $A=1000$
and $F_{a}=10^{12}$ GeV. The orange and blue regions are excluded
by the observation of the isocurvature perturbations and cold dark
matter abundance respectively.}
\label{Linde-A}
\end{figure}

\begin{figure}[H]
\centering\includegraphics[scale=0.5]{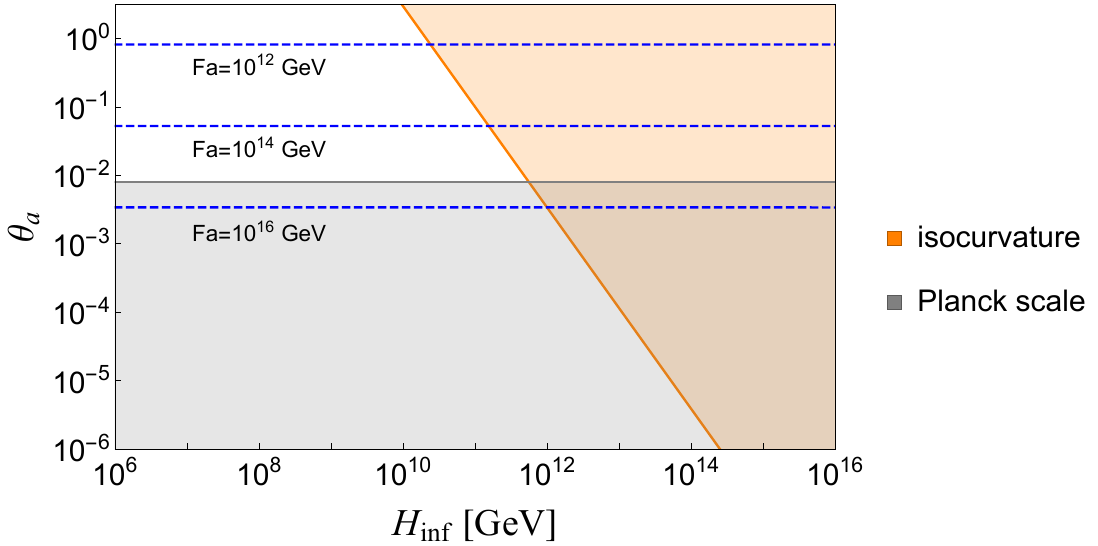}
\caption{Constraints on $\theta_a$ and $H_\text{inf}$ for $A=1000$ under the assumption
that the axion is the only component of cold dark matter. The gray region
is excluded by the condition that $Av=N\times AF_a \lesssim M_{{\rm pl}}$ for $N=1$. The blue dotted lines show $F_a=10^{12}$ GeV, $F_a=10^{14}$ GeV, and $F_a=10^{16}$ GeV.}
\label{Linde-A1000}
\end{figure}

In this case the isocurvature perturbation constraint on the Hubble
parameter during inflation $H_{\text{inf}}$ is shown as Fig.~\ref{Linde-A}
for $A=1000$ and $F_{a}=10^{12}\ {\rm GeV}$.
From the figure we find that the Hubble parameter $H_{{\rm inf}}$ can be as large as $10^{13}$~GeV if the misalignment angle is smaller than about $10^{-3}$.  
Furthermore if we assume that the axion accounts for all the dark
matter of the universe, the relation
\begin{align}
0.18\times\left[\theta_{a}^{2}+\left(\frac{H_{{\rm inf}}}{2\pi AF_{a}}\right)^{2}\right]\left(\frac{F_{a}}{10^{12}\text{GeV}}\right)^{1.19} & =0.12,\label{eq:axion=00003Dcdm}
\end{align}
holds. Using this equation and Eq.~(\ref{eq:iso}), we obtain the constraint
on $H_{{\rm inf}}$ and $\theta_{a}$ for $A=1000$ as shown in Fig.~\ref{Linde-A1000}. 
In the figure we also show the constraint (gray region) from the requirement that the field value during inflation should be less than $M_\text{pl}$, i.e. $Av \lesssim M_\text{pl}$.
We find that $H_{{\rm inf}}\lesssim10^{12}\ {\rm GeV}$ for $A=1000$
in the SUSY axion model without the domain wall problem. Thus the
axion can be the main component of the dark matter avoiding both domain
wall and isocurvautre perturbation problems for $H_{{\rm inf}}\lesssim10^{12}$
GeV in this model.

As mentioned before, with the present computer power it is difficult to examine whether domain walls are formed clearly for larger $A$.
However, results of $A\leq 1000$ imply that the fluctuations of the misalignment angle are not enough to produce domain walls.
This is because completely randam spatial distribution
of the axion after parametric resonance is needed for the domain wall
formation, but the situation is far from this even for $A=1000$ as
we can see from Fig.~\ref{fig:Fa_500}. Therefore we expect that the allowed parameter
region in Fig.~\ref{Linde-A1000} would be extended.

\section{Conclusion and discussion \label{sec:Conclusion-and-discussion}}


In this paper, we have investigated the isocurvature and domain wall
problems in the SUSY axion model. If the PQ field has a large field
value during inflation the isocuravture perturbations are suppressed.
However, after inflation the PQ field starts oscillation and produces
large fluctuations of the axion field through parametric resonance,
which may lead to the domain wall problem. We have performed the lattice
simulations to examine whether the domain wall problem arises. From the
result of the simulations, we have found that the domain walls are
not formed if the ratio $A$ of the PQ field value $|\Phi_{+}|$ during
inflation to the PQ breaking scale $v$ is less than $10^{3}$, i.e.
$A=|\Phi_{+}|_i/v\leq10^{3}$.
This relaxes the stringent constraint on the Hubble parameter during inflation $H_\text{inf}$ as $10^{13}$~GeV for $F_a =10^{12}$~GeV. 
Moreover, if the axion is dark matter, we obtain the constraint on $H_\text{inf}$ as $H_\text{inf} \lesssim 10^{12}$~GeV.  

We could not reach the definite conclusion on the domain wall problem
for larger $A$ because more machine power is needed to perform the
lattice simulations. However, for domain walls to be formed the spatial
distribution of the axion fields must be completely random, which seems
unlikely even for larger $A$ judging from the result of our simulations.
Therefore, we expect that larger $H_\text{inf}$ would be allowed. 

\section*{Acknowledgements}

This work was supported by JSPS KAKENHI Grant Nos.~17H01131 (M.K.)
and 17K05434 (M.K.), MEXT KAKENHI Grant No.~15H05889 (M.K.), and
World Premier International Research Center Initiative (WPI Initiative),
MEXT, Japan.


\bibliographystyle{apsrev4-1}
\bibliography{SUSY_axion}

\begin{thebibliography}{35}%
\makeatletter
\providecommand \@ifxundefined [1]{%
 \@ifx{#1\undefined}
}%
\providecommand \@ifnum [1]{%
 \ifnum #1\expandafter \@firstoftwo
 \else \expandafter \@secondoftwo
 \fi
}%
\providecommand \@ifx [1]{%
 \ifx #1\expandafter \@firstoftwo
 \else \expandafter \@secondoftwo
 \fi
}%
\providecommand \natexlab [1]{#1}%
\providecommand \enquote  [1]{``#1''}%
\providecommand \bibnamefont  [1]{#1}%
\providecommand \bibfnamefont [1]{#1}%
\providecommand \citenamefont [1]{#1}%
\providecommand \href@noop [0]{\@secondoftwo}%
\providecommand \href [0]{\begingroup \@sanitize@url \@href}%
\providecommand \@href[1]{\@@startlink{#1}\@@href}%
\providecommand \@@href[1]{\endgroup#1\@@endlink}%
\providecommand \@sanitize@url [0]{\catcode `\\12\catcode `\$12\catcode
  `\&12\catcode `\#12\catcode `\^12\catcode `\_12\catcode `\%12\relax}%
\providecommand \@@startlink[1]{}%
\providecommand \@@endlink[0]{}%
\providecommand \url  [0]{\begingroup\@sanitize@url \@url }%
\providecommand \@url [1]{\endgroup\@href {#1}{\urlprefix }}%
\providecommand \urlprefix  [0]{URL }%
\providecommand \Eprint [0]{\href }%
\providecommand \doibase [0]{http://dx.doi.org/}%
\providecommand \selectlanguage [0]{\@gobble}%
\providecommand \bibinfo  [0]{\@secondoftwo}%
\providecommand \bibfield  [0]{\@secondoftwo}%
\providecommand \translation [1]{[#1]}%
\providecommand \BibitemOpen [0]{}%
\providecommand \bibitemStop [0]{}%
\providecommand \bibitemNoStop [0]{.\EOS\space}%
\providecommand \EOS [0]{\spacefactor3000\relax}%
\providecommand \BibitemShut  [1]{\csname bibitem#1\endcsname}%
\let\auto@bib@innerbib\@empty
\bibitem [{\citenamefont {Weinberg}(1978)}]{Weinberg:1977ma}%
  \BibitemOpen
  \bibfield  {author} {\bibinfo {author} {\bibfnamefont {S.}~\bibnamefont
  {Weinberg}},\ }\href {\doibase 10.1103/PhysRevLett.40.223} {\bibfield
  {journal} {\bibinfo  {journal} {Phys. Rev. Lett.}\ }\textbf {\bibinfo
  {volume} {40}},\ \bibinfo {pages} {223} (\bibinfo {year} {1978})}\BibitemShut
  {NoStop}%
\bibitem [{\citenamefont {Wilczek}(1978)}]{Wilczek:1977pj}%
  \BibitemOpen
  \bibfield  {author} {\bibinfo {author} {\bibfnamefont {F.}~\bibnamefont
  {Wilczek}},\ }\href {\doibase 10.1103/PhysRevLett.40.279} {\bibfield
  {journal} {\bibinfo  {journal} {Phys. Rev. Lett.}\ }\textbf {\bibinfo
  {volume} {40}},\ \bibinfo {pages} {279} (\bibinfo {year} {1978})}\BibitemShut
  {NoStop}%
\bibitem [{\citenamefont {Peccei}\ and\ \citenamefont
  {Quinn}(1977{\natexlab{a}})}]{Peccei:1977ur}%
  \BibitemOpen
  \bibfield  {author} {\bibinfo {author} {\bibfnamefont {R.~D.}\ \bibnamefont
  {Peccei}}\ and\ \bibinfo {author} {\bibfnamefont {H.~R.}\ \bibnamefont
  {Quinn}},\ }\href {\doibase 10.1103/PhysRevD.16.1791} {\bibfield  {journal}
  {\bibinfo  {journal} {Phys. Rev.}\ }\textbf {\bibinfo {volume} {D16}},\
  \bibinfo {pages} {1791} (\bibinfo {year} {1977}{\natexlab{a}})}\BibitemShut
  {NoStop}%
\bibitem [{\citenamefont {Peccei}\ and\ \citenamefont
  {Quinn}(1977{\natexlab{b}})}]{Peccei:1977hh}%
  \BibitemOpen
  \bibfield  {author} {\bibinfo {author} {\bibfnamefont {R.~D.}\ \bibnamefont
  {Peccei}}\ and\ \bibinfo {author} {\bibfnamefont {H.~R.}\ \bibnamefont
  {Quinn}},\ }\href {\doibase 10.1103/PhysRevLett.38.1440} {\bibfield
  {journal} {\bibinfo  {journal} {Phys. Rev. Lett.}\ }\textbf {\bibinfo
  {volume} {38}},\ \bibinfo {pages} {1440} (\bibinfo {year}
  {1977}{\natexlab{b}})}\BibitemShut {NoStop}%
\bibitem [{\citenamefont {'t~Hooft}(1976)}]{tHooft:1976rip}%
  \BibitemOpen
  \bibfield  {author} {\bibinfo {author} {\bibfnamefont {G.}~\bibnamefont
  {'t~Hooft}},\ }\href {\doibase 10.1103/PhysRevLett.37.8} {\bibfield
  {journal} {\bibinfo  {journal} {Phys. Rev. Lett.}\ }\textbf {\bibinfo
  {volume} {37}},\ \bibinfo {pages} {8} (\bibinfo {year} {1976})}\BibitemShut
  {NoStop}%
\bibitem [{\citenamefont {Preskill}\ \emph {et~al.}(1983)\citenamefont
  {Preskill}, \citenamefont {Wise},\ and\ \citenamefont
  {Wilczek}}]{Preskill:1982cy}%
  \BibitemOpen
  \bibfield  {author} {\bibinfo {author} {\bibfnamefont {J.}~\bibnamefont
  {Preskill}}, \bibinfo {author} {\bibfnamefont {M.~B.}\ \bibnamefont {Wise}},
  \ and\ \bibinfo {author} {\bibfnamefont {F.}~\bibnamefont {Wilczek}},\ }\href
  {\doibase 10.1016/0370-2693(83)90637-8} {\bibfield  {journal} {\bibinfo
  {journal} {Phys. Lett.}\ }\textbf {\bibinfo {volume} {120B}},\ \bibinfo
  {pages} {127} (\bibinfo {year} {1983})}\BibitemShut {NoStop}%
\bibitem [{\citenamefont {Abbott}\ and\ \citenamefont
  {Sikivie}(1983)}]{Abbott:1982af}%
  \BibitemOpen
  \bibfield  {author} {\bibinfo {author} {\bibfnamefont {L.~F.}\ \bibnamefont
  {Abbott}}\ and\ \bibinfo {author} {\bibfnamefont {P.}~\bibnamefont
  {Sikivie}},\ }\href {\doibase 10.1016/0370-2693(83)90638-X} {\bibfield
  {journal} {\bibinfo  {journal} {Phys. Lett.}\ }\textbf {\bibinfo {volume}
  {120B}},\ \bibinfo {pages} {133} (\bibinfo {year} {1983})}\BibitemShut
  {NoStop}%
\bibitem [{\citenamefont {Dine}\ and\ \citenamefont
  {Fischler}(1983)}]{Dine:1982ah}%
  \BibitemOpen
  \bibfield  {author} {\bibinfo {author} {\bibfnamefont {M.}~\bibnamefont
  {Dine}}\ and\ \bibinfo {author} {\bibfnamefont {W.}~\bibnamefont
  {Fischler}},\ }\href {\doibase 10.1016/0370-2693(83)90639-1} {\bibfield
  {journal} {\bibinfo  {journal} {Phys. Lett.}\ }\textbf {\bibinfo {volume}
  {120B}},\ \bibinfo {pages} {137} (\bibinfo {year} {1983})}\BibitemShut
  {NoStop}%
\bibitem [{\citenamefont {Sikivie}(1982)}]{Sikivie:1982qv}%
  \BibitemOpen
  \bibfield  {author} {\bibinfo {author} {\bibfnamefont {P.}~\bibnamefont
  {Sikivie}},\ }\href {\doibase 10.1103/PhysRevLett.48.1156} {\bibfield
  {journal} {\bibinfo  {journal} {Phys. Rev. Lett.}\ }\textbf {\bibinfo
  {volume} {48}},\ \bibinfo {pages} {1156} (\bibinfo {year}
  {1982})}\BibitemShut {NoStop}%
\bibitem [{\citenamefont {Vilenkin}\ and\ \citenamefont
  {Everett}(1982)}]{Vilenkin:1982ks}%
  \BibitemOpen
  \bibfield  {author} {\bibinfo {author} {\bibfnamefont {A.}~\bibnamefont
  {Vilenkin}}\ and\ \bibinfo {author} {\bibfnamefont {A.~E.}\ \bibnamefont
  {Everett}},\ }\href {\doibase 10.1103/PhysRevLett.48.1867} {\bibfield
  {journal} {\bibinfo  {journal} {Phys. Rev. Lett.}\ }\textbf {\bibinfo
  {volume} {48}},\ \bibinfo {pages} {1867} (\bibinfo {year}
  {1982})}\BibitemShut {NoStop}%
\bibitem [{\citenamefont {Hiramatsu}\ \emph {et~al.}(2012)\citenamefont
  {Hiramatsu}, \citenamefont {Kawasaki}, \citenamefont {Saikawa},\ and\
  \citenamefont {Sekiguchi}}]{Hiramatsu:2012gg}%
  \BibitemOpen
  \bibfield  {author} {\bibinfo {author} {\bibfnamefont {T.}~\bibnamefont
  {Hiramatsu}}, \bibinfo {author} {\bibfnamefont {M.}~\bibnamefont {Kawasaki}},
  \bibinfo {author} {\bibfnamefont {K.}~\bibnamefont {Saikawa}}, \ and\
  \bibinfo {author} {\bibfnamefont {T.}~\bibnamefont {Sekiguchi}},\ }\href
  {\doibase 10.1103/PhysRevD.86.089902, 10.1103/PhysRevD.85.105020} {\bibfield
  {journal} {\bibinfo  {journal} {Phys. Rev.}\ }\textbf {\bibinfo {volume}
  {D85}},\ \bibinfo {pages} {105020} (\bibinfo {year} {2012})},\ \bibinfo
  {note} {[Erratum: Phys. Rev.D86,089902(2012)]},\ \Eprint
  {http://arxiv.org/abs/1202.5851} {arXiv:1202.5851 [hep-ph]} \BibitemShut
  {NoStop}%
\bibitem [{\citenamefont {Ryden}\ \emph {et~al.}(1989)\citenamefont {Ryden},
  \citenamefont {Press},\ and\ \citenamefont {Spergel}}]{Ryden:1989vj}%
  \BibitemOpen
  \bibfield  {author} {\bibinfo {author} {\bibfnamefont {B.~S.}\ \bibnamefont
  {Ryden}}, \bibinfo {author} {\bibfnamefont {W.~H.}\ \bibnamefont {Press}}, \
  and\ \bibinfo {author} {\bibfnamefont {D.~N.}\ \bibnamefont {Spergel}},\
  }\href@noop {} {\bibfield  {journal} {\bibinfo  {journal} {Submitted to:
  Astrophys. J.}\ } (\bibinfo {year} {1989})}\BibitemShut {NoStop}%
\bibitem [{\citenamefont {Axenides}\ \emph {et~al.}(1983)\citenamefont
  {Axenides}, \citenamefont {Brandenberger},\ and\ \citenamefont
  {Turner}}]{Axenides:1983hj}%
  \BibitemOpen
  \bibfield  {author} {\bibinfo {author} {\bibfnamefont {M.}~\bibnamefont
  {Axenides}}, \bibinfo {author} {\bibfnamefont {R.~H.}\ \bibnamefont
  {Brandenberger}}, \ and\ \bibinfo {author} {\bibfnamefont {M.~S.}\
  \bibnamefont {Turner}},\ }\href {\doibase 10.1016/0370-2693(83)90586-5}
  {\bibfield  {journal} {\bibinfo  {journal} {Phys. Lett.}\ }\textbf {\bibinfo
  {volume} {126B}},\ \bibinfo {pages} {178} (\bibinfo {year}
  {1983})}\BibitemShut {NoStop}%
\bibitem [{\citenamefont {Seckel}\ and\ \citenamefont
  {Turner}(1985)}]{Seckel:1985tj}%
  \BibitemOpen
  \bibfield  {author} {\bibinfo {author} {\bibfnamefont {D.}~\bibnamefont
  {Seckel}}\ and\ \bibinfo {author} {\bibfnamefont {M.~S.}\ \bibnamefont
  {Turner}},\ }\href {\doibase 10.1103/PhysRevD.32.3178} {\bibfield  {journal}
  {\bibinfo  {journal} {Phys. Rev.}\ }\textbf {\bibinfo {volume} {D32}},\
  \bibinfo {pages} {3178} (\bibinfo {year} {1985})}\BibitemShut {NoStop}%
\bibitem [{\citenamefont {Linde}(1985)}]{Linde:1985yf}%
  \BibitemOpen
  \bibfield  {author} {\bibinfo {author} {\bibfnamefont {A.~D.}\ \bibnamefont
  {Linde}},\ }\href {\doibase 10.1016/0370-2693(85)90436-8} {\bibfield
  {journal} {\bibinfo  {journal} {Phys. Lett.}\ }\textbf {\bibinfo {volume}
  {158B}},\ \bibinfo {pages} {375} (\bibinfo {year} {1985})}\BibitemShut
  {NoStop}%
\bibitem [{\citenamefont {Linde}\ and\ \citenamefont
  {Lyth}(1990)}]{Linde:1990yj}%
  \BibitemOpen
  \bibfield  {author} {\bibinfo {author} {\bibfnamefont {A.~D.}\ \bibnamefont
  {Linde}}\ and\ \bibinfo {author} {\bibfnamefont {D.~H.}\ \bibnamefont
  {Lyth}},\ }\href {\doibase 10.1016/0370-2693(90)90613-B} {\bibfield
  {journal} {\bibinfo  {journal} {Phys. Lett.}\ }\textbf {\bibinfo {volume}
  {B246}},\ \bibinfo {pages} {353} (\bibinfo {year} {1990})}\BibitemShut
  {NoStop}%
\bibitem [{\citenamefont {Turner}\ and\ \citenamefont
  {Wilczek}(1991)}]{Turner:1990uz}%
  \BibitemOpen
  \bibfield  {author} {\bibinfo {author} {\bibfnamefont {M.~S.}\ \bibnamefont
  {Turner}}\ and\ \bibinfo {author} {\bibfnamefont {F.}~\bibnamefont
  {Wilczek}},\ }\href {\doibase 10.1103/PhysRevLett.66.5} {\bibfield  {journal}
  {\bibinfo  {journal} {Phys. Rev. Lett.}\ }\textbf {\bibinfo {volume} {66}},\
  \bibinfo {pages} {5} (\bibinfo {year} {1991})}\BibitemShut {NoStop}%
\bibitem [{\citenamefont {Lyth}(1992)}]{Lyth:1991ub}%
  \BibitemOpen
  \bibfield  {author} {\bibinfo {author} {\bibfnamefont {D.~H.}\ \bibnamefont
  {Lyth}},\ }\href {\doibase 10.1103/PhysRevD.45.3394} {\bibfield  {journal}
  {\bibinfo  {journal} {Phys. Rev.}\ }\textbf {\bibinfo {volume} {D45}},\
  \bibinfo {pages} {3394} (\bibinfo {year} {1992})}\BibitemShut {NoStop}%
\bibitem [{\citenamefont {Ade}\ \emph {et~al.}(2016)\citenamefont {Ade} \emph
  {et~al.}}]{Ade:2015lrj}%
  \BibitemOpen
  \bibfield  {author} {\bibinfo {author} {\bibfnamefont {P.~A.~R.}\
  \bibnamefont {Ade}} \emph {et~al.} (\bibinfo {collaboration} {Planck}),\
  }\href {\doibase 10.1051/0004-6361/201525898} {\bibfield  {journal} {\bibinfo
   {journal} {Astron. Astrophys.}\ }\textbf {\bibinfo {volume} {594}},\
  \bibinfo {pages} {A20} (\bibinfo {year} {2016})},\ \Eprint
  {http://arxiv.org/abs/1502.02114} {arXiv:1502.02114 [astro-ph.CO]}
  \BibitemShut {NoStop}%
\bibitem [{\citenamefont {Linde}(1991)}]{Linde:1991km}%
  \BibitemOpen
  \bibfield  {author} {\bibinfo {author} {\bibfnamefont {A.~D.}\ \bibnamefont
  {Linde}},\ }\href {\doibase 10.1016/0370-2693(91)90130-I} {\bibfield
  {journal} {\bibinfo  {journal} {Phys. Lett.}\ }\textbf {\bibinfo {volume}
  {B259}},\ \bibinfo {pages} {38} (\bibinfo {year} {1991})}\BibitemShut
  {NoStop}%
\bibitem [{\citenamefont {Kofman}\ \emph {et~al.}(1996)\citenamefont {Kofman},
  \citenamefont {Linde},\ and\ \citenamefont {Starobinsky}}]{Kofman:1995fi}%
  \BibitemOpen
  \bibfield  {author} {\bibinfo {author} {\bibfnamefont {L.}~\bibnamefont
  {Kofman}}, \bibinfo {author} {\bibfnamefont {A.~D.}\ \bibnamefont {Linde}}, \
  and\ \bibinfo {author} {\bibfnamefont {A.~A.}\ \bibnamefont {Starobinsky}},\
  }\href {\doibase 10.1103/PhysRevLett.76.1011} {\bibfield  {journal} {\bibinfo
   {journal} {Phys. Rev. Lett.}\ }\textbf {\bibinfo {volume} {76}},\ \bibinfo
  {pages} {1011} (\bibinfo {year} {1996})},\ \Eprint
  {http://arxiv.org/abs/hep-th/9510119} {arXiv:hep-th/9510119 [hep-th]}
  \BibitemShut {NoStop}%
\bibitem [{\citenamefont {Kofman}\ \emph {et~al.}(1997)\citenamefont {Kofman},
  \citenamefont {Linde},\ and\ \citenamefont {Starobinsky}}]{Kofman:1997yn}%
  \BibitemOpen
  \bibfield  {author} {\bibinfo {author} {\bibfnamefont {L.}~\bibnamefont
  {Kofman}}, \bibinfo {author} {\bibfnamefont {A.~D.}\ \bibnamefont {Linde}}, \
  and\ \bibinfo {author} {\bibfnamefont {A.~A.}\ \bibnamefont {Starobinsky}},\
  }\href {\doibase 10.1103/PhysRevD.56.3258} {\bibfield  {journal} {\bibinfo
  {journal} {Phys. Rev.}\ }\textbf {\bibinfo {volume} {D56}},\ \bibinfo {pages}
  {3258} (\bibinfo {year} {1997})},\ \Eprint
  {http://arxiv.org/abs/hep-ph/9704452} {arXiv:hep-ph/9704452 [hep-ph]}
  \BibitemShut {NoStop}%
\bibitem [{\citenamefont {Shtanov}\ \emph {et~al.}(1995)\citenamefont
  {Shtanov}, \citenamefont {Traschen},\ and\ \citenamefont
  {Brandenberger}}]{Shtanov:1994ce}%
  \BibitemOpen
  \bibfield  {author} {\bibinfo {author} {\bibfnamefont {Y.}~\bibnamefont
  {Shtanov}}, \bibinfo {author} {\bibfnamefont {J.~H.}\ \bibnamefont
  {Traschen}}, \ and\ \bibinfo {author} {\bibfnamefont {R.~H.}\ \bibnamefont
  {Brandenberger}},\ }\href {\doibase 10.1103/PhysRevD.51.5438} {\bibfield
  {journal} {\bibinfo  {journal} {Phys. Rev.}\ }\textbf {\bibinfo {volume}
  {D51}},\ \bibinfo {pages} {5438} (\bibinfo {year} {1995})},\ \Eprint
  {http://arxiv.org/abs/hep-ph/9407247} {arXiv:hep-ph/9407247 [hep-ph]}
  \BibitemShut {NoStop}%
\bibitem [{\citenamefont {Tkachev}\ \emph {et~al.}(1998)\citenamefont
  {Tkachev}, \citenamefont {Khlebnikov}, \citenamefont {Kofman},\ and\
  \citenamefont {Linde}}]{Tkachev:1998dc}%
  \BibitemOpen
  \bibfield  {author} {\bibinfo {author} {\bibfnamefont {I.}~\bibnamefont
  {Tkachev}}, \bibinfo {author} {\bibfnamefont {S.}~\bibnamefont {Khlebnikov}},
  \bibinfo {author} {\bibfnamefont {L.}~\bibnamefont {Kofman}}, \ and\ \bibinfo
  {author} {\bibfnamefont {A.~D.}\ \bibnamefont {Linde}},\ }\href {\doibase
  10.1016/S0370-2693(98)01094-6} {\bibfield  {journal} {\bibinfo  {journal}
  {Phys. Lett.}\ }\textbf {\bibinfo {volume} {B440}},\ \bibinfo {pages} {262}
  (\bibinfo {year} {1998})},\ \Eprint {http://arxiv.org/abs/hep-ph/9805209}
  {arXiv:hep-ph/9805209 [hep-ph]} \BibitemShut {NoStop}%
\bibitem [{\citenamefont {Kasuya}\ and\ \citenamefont
  {Kawasaki}(1998)}]{Kasuya:1998td}%
  \BibitemOpen
  \bibfield  {author} {\bibinfo {author} {\bibfnamefont {S.}~\bibnamefont
  {Kasuya}}\ and\ \bibinfo {author} {\bibfnamefont {M.}~\bibnamefont
  {Kawasaki}},\ }\href {\doibase 10.1103/PhysRevD.58.083516} {\bibfield
  {journal} {\bibinfo  {journal} {Phys. Rev.}\ }\textbf {\bibinfo {volume}
  {D58}},\ \bibinfo {pages} {083516} (\bibinfo {year} {1998})},\ \Eprint
  {http://arxiv.org/abs/hep-ph/9804429} {arXiv:hep-ph/9804429 [hep-ph]}
  \BibitemShut {NoStop}%
\bibitem [{\citenamefont {Kasuya}\ and\ \citenamefont
  {Kawasaki}(2000)}]{Kasuya:1999hy}%
  \BibitemOpen
  \bibfield  {author} {\bibinfo {author} {\bibfnamefont {S.}~\bibnamefont
  {Kasuya}}\ and\ \bibinfo {author} {\bibfnamefont {M.}~\bibnamefont
  {Kawasaki}},\ }\href {\doibase 10.1103/PhysRevD.61.083510} {\bibfield
  {journal} {\bibinfo  {journal} {Phys. Rev.}\ }\textbf {\bibinfo {volume}
  {D61}},\ \bibinfo {pages} {083510} (\bibinfo {year} {2000})},\ \Eprint
  {http://arxiv.org/abs/hep-ph/9903324} {arXiv:hep-ph/9903324 [hep-ph]}
  \BibitemShut {NoStop}%
\bibitem [{\citenamefont {Kawasaki}\ \emph {et~al.}(2013)\citenamefont
  {Kawasaki}, \citenamefont {Yanagida},\ and\ \citenamefont
  {Yoshino}}]{Kawasaki:2013iha}%
  \BibitemOpen
  \bibfield  {author} {\bibinfo {author} {\bibfnamefont {M.}~\bibnamefont
  {Kawasaki}}, \bibinfo {author} {\bibfnamefont {T.~T.}\ \bibnamefont
  {Yanagida}}, \ and\ \bibinfo {author} {\bibfnamefont {K.}~\bibnamefont
  {Yoshino}},\ }\href {\doibase 10.1088/1475-7516/2013/11/030} {\bibfield
  {journal} {\bibinfo  {journal} {JCAP}\ }\textbf {\bibinfo {volume} {1311}},\
  \bibinfo {pages} {030} (\bibinfo {year} {2013})},\ \Eprint
  {http://arxiv.org/abs/1305.5338} {arXiv:1305.5338 [hep-ph]} \BibitemShut
  {NoStop}%
\bibitem [{\citenamefont {Kasuya}\ \emph {et~al.}(1997)\citenamefont {Kasuya},
  \citenamefont {Kawasaki},\ and\ \citenamefont {Yanagida}}]{Kasuya:1996ns}%
  \BibitemOpen
  \bibfield  {author} {\bibinfo {author} {\bibfnamefont {S.}~\bibnamefont
  {Kasuya}}, \bibinfo {author} {\bibfnamefont {M.}~\bibnamefont {Kawasaki}}, \
  and\ \bibinfo {author} {\bibfnamefont {T.}~\bibnamefont {Yanagida}},\ }\href
  {\doibase 10.1016/S0370-2693(97)00809-5} {\bibfield  {journal} {\bibinfo
  {journal} {Phys. Lett.}\ }\textbf {\bibinfo {volume} {B409}},\ \bibinfo
  {pages} {94} (\bibinfo {year} {1997})},\ \Eprint
  {http://arxiv.org/abs/hep-ph/9608405} {arXiv:hep-ph/9608405 [hep-ph]}
  \BibitemShut {NoStop}%
\bibitem [{\citenamefont {Dine}\ \emph {et~al.}(1995)\citenamefont {Dine},
  \citenamefont {Randall},\ and\ \citenamefont {Thomas}}]{Dine:1995uk}%
  \BibitemOpen
  \bibfield  {author} {\bibinfo {author} {\bibfnamefont {M.}~\bibnamefont
  {Dine}}, \bibinfo {author} {\bibfnamefont {L.}~\bibnamefont {Randall}}, \
  and\ \bibinfo {author} {\bibfnamefont {S.~D.}\ \bibnamefont {Thomas}},\
  }\href {\doibase 10.1103/PhysRevLett.75.398} {\bibfield  {journal} {\bibinfo
  {journal} {Phys. Rev. Lett.}\ }\textbf {\bibinfo {volume} {75}},\ \bibinfo
  {pages} {398} (\bibinfo {year} {1995})},\ \Eprint
  {http://arxiv.org/abs/hep-ph/9503303} {arXiv:hep-ph/9503303 [hep-ph]}
  \BibitemShut {NoStop}%
\bibitem [{\citenamefont {Gaillard}\ \emph {et~al.}(1995)\citenamefont
  {Gaillard}, \citenamefont {Murayama},\ and\ \citenamefont
  {Olive}}]{Gaillard:1995az}%
  \BibitemOpen
  \bibfield  {author} {\bibinfo {author} {\bibfnamefont {M.~K.}\ \bibnamefont
  {Gaillard}}, \bibinfo {author} {\bibfnamefont {H.}~\bibnamefont {Murayama}},
  \ and\ \bibinfo {author} {\bibfnamefont {K.~A.}\ \bibnamefont {Olive}},\
  }\href {\doibase 10.1016/0370-2693(95)00773-E} {\bibfield  {journal}
  {\bibinfo  {journal} {Phys. Lett.}\ }\textbf {\bibinfo {volume} {B355}},\
  \bibinfo {pages} {71} (\bibinfo {year} {1995})},\ \Eprint
  {http://arxiv.org/abs/hep-ph/9504307} {arXiv:hep-ph/9504307 [hep-ph]}
  \BibitemShut {NoStop}%
\bibitem [{\citenamefont {Kim}(1979)}]{Kim:1979if}%
  \BibitemOpen
  \bibfield  {author} {\bibinfo {author} {\bibfnamefont {J.~E.}\ \bibnamefont
  {Kim}},\ }\href {\doibase 10.1103/PhysRevLett.43.103} {\bibfield  {journal}
  {\bibinfo  {journal} {Phys. Rev. Lett.}\ }\textbf {\bibinfo {volume} {43}},\
  \bibinfo {pages} {103} (\bibinfo {year} {1979})}\BibitemShut {NoStop}%
\bibitem [{\citenamefont {Shifman}\ \emph {et~al.}(1980)\citenamefont
  {Shifman}, \citenamefont {Vainshtein},\ and\ \citenamefont
  {Zakharov}}]{Shifman:1979if}%
  \BibitemOpen
  \bibfield  {author} {\bibinfo {author} {\bibfnamefont {M.~A.}\ \bibnamefont
  {Shifman}}, \bibinfo {author} {\bibfnamefont {A.~I.}\ \bibnamefont
  {Vainshtein}}, \ and\ \bibinfo {author} {\bibfnamefont {V.~I.}\ \bibnamefont
  {Zakharov}},\ }\href {\doibase 10.1016/0550-3213(80)90209-6} {\bibfield
  {journal} {\bibinfo  {journal} {Nucl. Phys.}\ }\textbf {\bibinfo {volume}
  {B166}},\ \bibinfo {pages} {493} (\bibinfo {year} {1980})}\BibitemShut
  {NoStop}%
\bibitem [{\citenamefont {Dine}\ \emph {et~al.}(1981)\citenamefont {Dine},
  \citenamefont {Fischler},\ and\ \citenamefont {Srednicki}}]{Dine:1981rt}%
  \BibitemOpen
  \bibfield  {author} {\bibinfo {author} {\bibfnamefont {M.}~\bibnamefont
  {Dine}}, \bibinfo {author} {\bibfnamefont {W.}~\bibnamefont {Fischler}}, \
  and\ \bibinfo {author} {\bibfnamefont {M.}~\bibnamefont {Srednicki}},\ }\href
  {\doibase 10.1016/0370-2693(81)90590-6} {\bibfield  {journal} {\bibinfo
  {journal} {Phys. Lett.}\ }\textbf {\bibinfo {volume} {104B}},\ \bibinfo
  {pages} {199} (\bibinfo {year} {1981})}\BibitemShut {NoStop}%
\bibitem [{\citenamefont {Zhitnitsky}(1980)}]{Zhitnitsky:1980tq}%
  \BibitemOpen
  \bibfield  {author} {\bibinfo {author} {\bibfnamefont {A.~R.}\ \bibnamefont
  {Zhitnitsky}},\ }\href@noop {} {\bibfield  {journal} {\bibinfo  {journal}
  {Sov. J. Nucl. Phys.}\ }\textbf {\bibinfo {volume} {31}},\ \bibinfo {pages}
  {260} (\bibinfo {year} {1980})},\ \bibinfo {note} {[Yad.
  Fiz.31,497(1980)]}\BibitemShut {NoStop}%
\bibitem [{\citenamefont {Kim}\ and\ \citenamefont
  {Nilles}(1984)}]{Kim:1983dt}%
  \BibitemOpen
  \bibfield  {author} {\bibinfo {author} {\bibfnamefont {J.~E.}\ \bibnamefont
  {Kim}}\ and\ \bibinfo {author} {\bibfnamefont {H.~P.}\ \bibnamefont
  {Nilles}},\ }\href {\doibase 10.1016/0370-2693(84)91890-2} {\bibfield
  {journal} {\bibinfo  {journal} {Phys. Lett.}\ }\textbf {\bibinfo {volume}
  {138B}},\ \bibinfo {pages} {150} (\bibinfo {year} {1984})}\BibitemShut
  {NoStop}%
\end{thebibliography}%

\end{document}